\documentclass[preprint,12pt]{elsarticle}
\usepackage{multirow}
\usepackage{booktabs}
\usepackage{bm}% bold math

\usepackage{amssymb}
\usepackage{graphicx}
\usepackage[colorlinks,linkcolor=blue,anchorcolor=blue,citecolor=blue]{hyperref}

\biboptions{numbers,sort&compress}

\journal{}

\begin{document}

\begin{frontmatter}

%% Title, authors and addresses

%% use the tnoteref command within \title for footnotes;
%% use the tnotetext command for the associated footnote;
%% use the fnref command within \author or \address for footnotes;
%% use the fntext command for the associated footnote;
%% use the corref command within \author for corresponding author footnotes;
%% use the cortext command for the associated footnote;
%% use the ead command for the email address,
%% and the form \ead[url] for the home page:
%%
%% \title{Title\tnoteref{label1}}
%% \tnotetext[label1]{}
%% \author{Name\corref{cor1}\fnref{label2}}
%% \ead{email address}
%% \ead[url]{home page}
%% \fntext[label2]{}
%% \cortext[cor1]{}
%% \address{Address\fnref{label3}}
%% \fntext[label3]{}

\title{Identification of highly susceptible individuals in complex networks }

%% use optional labels to link authors explicitly to addresses:
%% \author[label1,label2]{<author name>}
%% \address[label1]{<address>}
%% \address[label2]{<address>}
\author[bh]{Shaoting Tang}
\ead{tangshaoting@buaa.edu.cn}
\author[bh]{Xian Teng}
\author[bh]{Sen Pei}
\ead{peisen@buaa.edu.cn}
\author[bh]{Shu Yan}
\author[bh]{Zhiming Zheng}
\ead{zzheng@pku.edu.cn}

\address[bh]{LMIB and School of Mathematics and Systems Science, Beihang University, Beijing, China}

\begin{abstract}
%% Text of abstract
Identifying highly susceptible individuals in spreading processes is of great significance in controlling outbreaks. In this paper, we explore the susceptibility of people in susceptible-infectious-recovered (SIR) and rumor spreading dynamics. We first study the impact of community structure on people's susceptibility. Despite that the community structure can reduce the infected population given same infection rates, it will not deterministically affect nodes' susceptibility. We find the susceptibility of individuals is sensitive to the choice of spreading dynamics. For SIR spreading, since the susceptibility is highly correlated to nodes' influence, the topological indicator $k$-shell can better identify highly susceptible individuals, outperforming degree, betweenness centrality and PageRank. In contrast, in rumor spreading model, where nodes' susceptibility and influence have no clear correlation, degree performs the best among considered topological measures. Our finding highlights the significance of both topological features and spreading mechanisms in identifying highly susceptible population.
\end{abstract}

\begin{keyword}
%% keywords here, in the form: keyword \sep keyword

Epidemic spreading \sep Susceptibility \sep Complex networks

%% MSC codes here, in the form: \MSC code \sep code
%% or \MSC[2008] code \sep code (2000 is the default)

\end{keyword}

\end{frontmatter}

%%
%% Start line numbering here if you want
%%
% \linenumbers

%% main text
\section{Introduction}

Developing efficient methods to prevent epidemic outbreaks or accelerate information dissemination is the ultimate goal of research on spreading dynamics across various domains \cite{Keeling2008,Brockmann2013,Maksim2010,Albert2000,Reis2014,Pei2012,Boguna2002,Moreno2004,Borge2012,Pei2012a,Hu2014,Teng2014,Pastor2001,Newman2002,Yan2013}. For diseases and information that spread through social networks, the structure of underlying contact network can greatly affect the spreading processes \cite{Pastor2001,Newman2002,Yan2013}. Previous studies have inspected the role of topology in epidemic threshold and critical behavior. For example, in uncorrelated networks, the epidemic threshold is $\lambda = \langle k \rangle / (\langle k^2 \rangle-\langle k \rangle)$, where $\langle k \rangle$ and $\langle k^2 \rangle$ are the first and second moments of the degree distribution \cite{Newman2002}. Later on, in view of the wide occurrence of intrinsic heterogeneous complex networks in real world, researchers started to investigate the effect of microstructure on spreading dynamics in more detailed perspectives: the community structure appearing in social networks \cite{Girvan2002}, $k$-shell decomposition of networks for identification of influential spreaders \cite{Maksim2010}, link salience skeleton correlated with the frequency of a link's appearance in infection hierarchies \cite{Daniel2011}, as well as weak ties which is significant for information dissemination \cite{Onnela2006}.

Due to the existence of complex microstructure, individuals with different topological features should play distinct roles during the spreading processes. Previous studies mainly focus on the spreading ability of individuals. Many topological measures are employed to identify influential spreaders in networks, such as degree \cite{Albert2000,Pastor2001}, betweenness \cite{Freeman1979}, $k$-shell index \cite{Maksim2010,Zeng2013}, PageRank \cite{Brin1998,Lu2011}, etc. In fact, individuals situated in different positions of the contact network will also vary in susceptibility to epidemics. Identifying the highly susceptible individuals in contact networks is of great significance in controlling epidemic outbreaks. Therefore, it is indispensable to explore the factors that affect people's susceptibility.

In this paper, we study the susceptibility of individuals in SIR and rumor spreading dynamics through extensive simulations in real-world social networks. First, we explore the impact of community structure on people's susceptibility. Even though the community structure can diminish the infected population and slow down the spreading processes, it is not the pivotal factor affecting nodes' susceptibility. Although the location of epidemic source has an impact on the precise infected probability of each single person, there exist a group of nodes that can always get infected with relatively high probabilities no matter where the spreading originates. We are particularly interested in these populations and consider them as highly susceptible individuals. To quantitatively depict the property of these people, we define the susceptibility of each individual as the average probability to be infected by a randomly chosen spreading source. Then we further explore the topological properties of the highly susceptibility individuals. By examining their topological traits including degree, $k$-shell, betweenness centrality and PageRank, we find the susceptibility of individuals is sensitive to the choice of spreading dynamics. For SIR spreading, the susceptibility is highly correlated to nodes' influence. Therefore, nodes located in the core region of networks are more likely to be infected. However, in rumor spreading model, degree can better identify highly susceptible individuals. Our results indicate that it is necessary to acquire the information of spreading mechanism to better locate highly susceptible population in practice.

\section{Impact of community structure on spreading processes}

Most of social networks in reality have notable community structure. We first explore the impact of community structure on individuals' probabilities to be infected in spreading processes. We examine the epidemic flows in two social networks with community structure: (1) the friendship network between users of the social network Facebook (Facebook) \cite{Leskovec2012}, (2) the Enron email communication network (Enron), in which nodes are email addresses and edges represent email communications \cite{Klimt2004}. These social networks have been used in previous studies. Topological statistics of networks are shown in Table \ref{tb}. In our following study, we treat all these networks as undirected.

\begin{table}
\caption{Statistics of social networks. We display the node number $N$, link number $L$, and average degree $\langle k\rangle$ for Facebook and Enron social networks. The critical infection rate of SIR model $\beta_c$ is calculated by $\langle k \rangle / (\langle k^2 \rangle-\langle k \rangle)$. $\beta_S$ and $\beta_R$ are adopted infection rates in SIR and rumor spreading simulations respectively. }
\label{tb}
\begin{center}
\begin{tabular}{ccccccc}
%\hline
\multirow{2}{*}{Network} &
\multicolumn{3}{c}{Topological statistics} &
\multirow{2}{*}{$\beta_c$} &
\multirow{2}{*}{$\beta_S$} &
\multirow{2}{*}{$\beta_R$}\\
\cline{2-4}
  & $N$ & $L$ & $\langle k\rangle$ &  \\
\toprule[1pt]
\multirow{1}{*}{Facebook} & $4,039$ & $88,234$ & $43.7$ & $0.0095$ & $0.03$ & $0.3$\\
\multirow{1}{*}{Enron} & $36,692$ & $367,662$ & $20.0$ & $0.0072$ & $0.02$ & $0.3$\\
\bottomrule[1pt]
\end{tabular}
\end{center}
\end{table}

\begin{figure}
\begin{center}
\includegraphics[width=0.8\columnwidth]{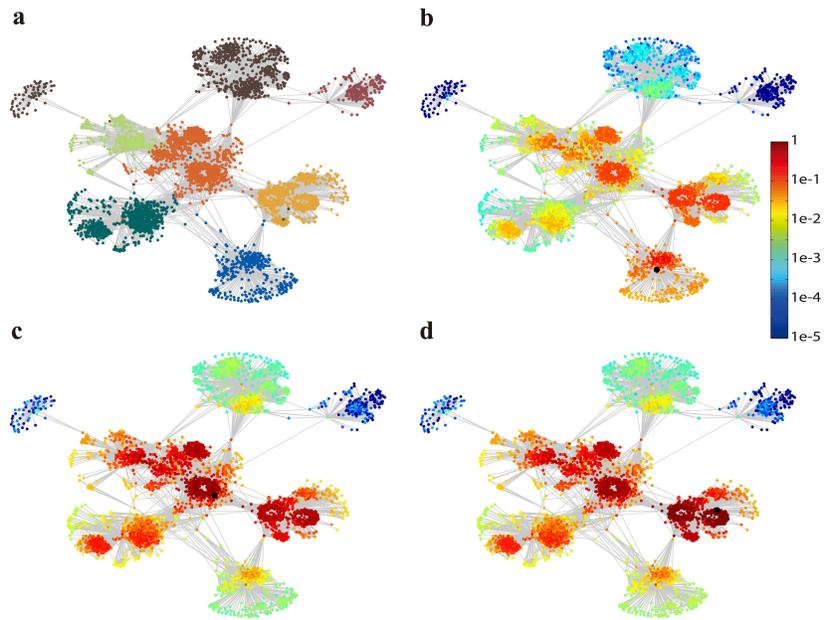} \caption{SIR epidemic spreading in Facebook social network. The infection rate is set as $\beta_S=0.03$. (a), The layout of community structure for Facebook social network. Eight communities are distinguished by different colors. (b, c, d), Distributions of infected probability for spreading sources in different communities. The solid black dots represent outbreak origins. The infected probability is obtained by averaging over $10^5$ SIR realizations. The colors indicate the logarithmic values (base 10) of infected probability for corresponding individuals.}
\label{fig1}
\end{center}
\end{figure}

The community structure is extracted by a fast heuristic method based on modularity optimization \cite{Blondel2008}. The layout of community structure for Facebook social network is shown with different colors in Fig. \ref{fig1}(a). To simulate spreading processes among populations, we first apply susceptible-infectious-recovered (SIR) \cite{Hethcote2000,Yan2014} model on the above social networks. In SIR model, the population can be classified into three possible states: susceptible (S), infected (I), or recovered (R). Starting from a single epidemic source node, at each time step, the infected individuals (I) would infect their susceptible neighbors (S) with probability $\beta_S$ and then become recovered (R) and gain permanent immunity with probability $\lambda$. Such process continues until there are no infected individuals in the system. In our simulations, without loss of generality, we set $\lambda=1$. Previous studies have adopted this model to describe diffusion of contagious diseases. For information dissemination, recent works have shown that, although some information spreading processes are greatly affected by human-related factors \cite{Centola2010,Centola2011,Karsai2014,Muchnik2013,Li2014}, a few information memes showing viral behavior can also spread like infectious diseases \cite{Weng2013}. Hence the SIR model is applicable not only to epidemics but also to those information memes which are not seriously affected by human-related factors.

In simulations, the selection of infection rate $\beta_S$ could significantly affect the spreading outcomes. Below the critical value $\beta_c=\langle k \rangle / (\langle k^2 \rangle-\langle k \rangle)$, the proportion of infected individuals will vanish in the limit of a very large population. For $\beta_S$ much larger than $\beta_c$, the epidemic will infect almost all the people in the network. In our study, we use relatively small values of $\beta_S$ to guarantee medium scale infection coverage. Otherwise most of the nodes will be infected with high probabilities, making their susceptibility indistinguishable. In Table \ref{tb}, we report the critical infection rate $\beta_c$ and adopted value $\beta_S$ in our simulations. The infection rate $\beta_S$ is slightly above its corresponding critical value $\beta_c$ to generate moderate coverage.

For Facebook network, we select three nodes in different communities as spreading sources and apply $10^5$ realizations of SIR model. The infected probability of each individual is approximated by the frequency of getting infected in these realizations. In Fig.\ref{fig1}(b-d), we display the distribution of infected probability for Facebook social network. The black dots represent outbreak origins and the infected probability is indicated by colors. Clearly, the epidemic permeates to several communities. Moreover, during all the epidemic spreading, there exist a group of individuals retaining relatively high infected probabilities, regardless of the positions of spreading sources.

\begin{figure}
\begin{center}
\includegraphics[width=0.8\columnwidth]{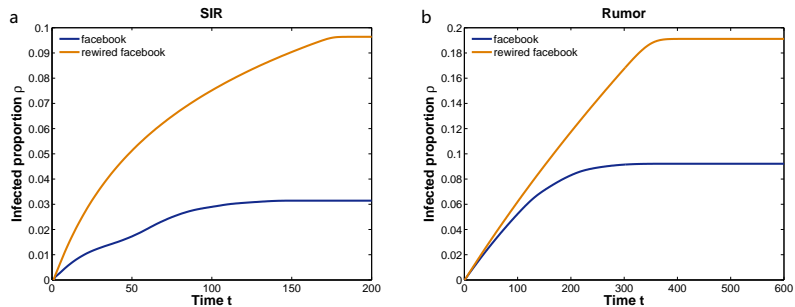} \caption{Community structure will reduce the infected population and slow down the spreading processes. For Facebook and rewired Facebook social networks, we perform SIR model with $\beta_S=0.03$. The evolution of infected proportion $\rho$ is displayed in (a). In (b), we present the infected proportion $\rho$ versus time $t$ for rumor spreading model with $\beta_R=0.3$.}
\label{fig2}
\end{center}
\end{figure}

In order to check the effect of community structure on spreading processes, we compare the evolution of infected proportion $\rho$ in networks with and without community structure. For Facebook network, we rewire the links to destroy the community structure. Specifically, we randomly select two links $(A, B)$ and $(C, D)$, and then swap the end nodes to $(A, D)$ and $(B, C)$. This procedure breaks down the communities but preserves the degree distribution. In our simulations, we perform $10^6$ rewire operations. The result of SIR model is shown in Fig.\ref{fig2}(a). Under same conditions, the rewired Facebook network has a much higher infected proportion, which implies the community structure restricts the spreading processes to some extent. Meanwhile, it will take a longer time for Facebook network with community structure to achieve a given infected proportion $\rho$. Therefore, the community structure will slow down the speed of SIR spreading.

To examine whether our finding is applicable to other spreading dynamics, we perform the same analysis on rumor spreading model \cite{Moreno2004}. In rumor spreading model, each individual can be in three possible states: the spreader (S), ignorant (I), and stifler (R). Spreaders represent nodes that are aware of the rumor and are willing to transmit it. Ignorant people are individuals unaware of the rumor. Stiflers stand for those that already know the rumor but are not willing to spread it anymore. In each time step, the spreaders contact one of their neighbors randomly and turn the ignorant ones into spreaders with probability $\beta_R$. If the spreaders encounter spreaders or stiflers, they will turn to stiflers with probability $\lambda$. We also set $\lambda=1$ in our simulations. We obtain similar results in Fig.\ref{fig2}(b).

\section{Susceptibility of individuals in spreading processes}

During spreading processes, the exact values of infected probability depends on the choice of spreading sources. But in real-world applications, usually we have no information about the locations of spreading origins, which makes it impractical and intractable to obtain individuals' exact infected probability during an epidemic outbreak. In spite of that, there exist a group of nodes exhibiting relatively high infected probability regardless of the position of source (see Fig.\ref{fig1}(b-d)). In our work, we will focus on these persons who would be infected easily from a randomly chosen spreading origin. We define the source-independent susceptibility $S_i$ of a given individual $i$ as the average infected probability over epidemics starting from all the nodes in the network:
\begin{equation}
S_i = \frac{1}{N}\sum_{j=1}^{N}\frac{\sum_{k=1}^{T}\pi_{ij}^{k}}{T} = \frac{1}{N}\sum_{j=1}^{N}p_{ij},
\label{Si}
\end{equation}
where $N$ is the number of nodes in the network, and $T$ is the number of spreading realizations originating from each node. In our
research, we set $T = 10^4$. Besides, $\pi_{ij}^{k}$ denotes whether node $i$ gets infected ($\pi_{ij}^k = 1$) or not ($\pi_{ij}^k = 0$) during the $k^{th}$ spreading realization originating from node $j$. Accordingly, $p_{ij} = \sum_{k=1}^{T}\pi_{ij}^{k}/T$ represents the average infected probability of node $i$ during the spreading from node $j$. We should note that this definition of susceptibility is different from that in \cite{Smilkov2014}, where the susceptibility is defined as the nature of individuals, such as immune system, physical condition or learning ability, etc. According to our definition, on average, nodes with higher susceptibility $S_i$ are more
likely to be infected during an epidemic wherever the contagion starts.

For SIR spreading model, we can calculate the susceptibility with the standard set of equations of SIR dynamics. For a network with $N$ nodes, the topology is recorded by the adjacency matrix $\bm{A}=\{a_{ij}\}_{N\times N}$, where $a_{ij}=1$ if node $i$ and $j$ are connected, and $a_{ij}=0$ otherwise. We denote the probability of node $i$ being in susceptible, infected and recovered state at time $t$ as $p_{i}^S(t)$, $p_{i}^I(t)$ and $p_{i}^R(t)$ respectively.  For networks without too many short loops, we assume the neighbors of one node can independently get infected at time $t$. Then the evolution of dynamics follows
\begin{eqnarray}
% \nonumber to remove numbering (before each equation)
  \frac{dp_{i}^S(t)}{dt}&=&-[1-\prod_{j=1}^N(1-\beta_S a_{ij}p_{j}^I(t)p_{i}^S(t))], \\
  \frac{dp_{i}^I(t)}{dt}&=&[1-\prod_{j=1}^N(1-\beta_S a_{ij}p_{j}^I(t)p_{i}^S(t))]-p_{i}^I(t), \\
  \frac{dp_{i}^R(t)}{dt}&=&p_{i}^I(t).
\end{eqnarray}
Here $a_{ij}$ is the element of adjacency matrix $\bm{A}$. In case of small infection rate $\beta_S$, the term $[1-\prod_{j=1}^N(1-\beta_S a_{ij}p_{j}^I(t)p_{i}^S(t))]$ can be approximated by $\beta_S p_{i}^S(t)\sum_{j=1}^N a_{ij}p_{j}^S(t)$. In fact, there are only two independent equations since $p_{i}^S(t)+p_{i}^I(t)+p_{i}^R(t)=1$. We discretize the above equations and obtain two independent ones
\begin{eqnarray}
% \nonumber to remove numbering (before each equation)
  p_{i}^I(t+1)&=&\beta_S(1-p_{i}^I(t)-p_{i}^R(t))\sum_{j=1}^N a_{ij}p_{j}^I(t), \\
  p_{i}^R(t+1)&=&p_{i}^R(t)+p_{i}^I(t).
\end{eqnarray}

For simplicity, the above equations can be transformed into a matrix form. Denote $\bm{p^I}(t)$ as the vector $(p_{0}^I(t),\cdots,p_{N}^I(t))^T$, $\bm{p^R}(t)$ as the vector $(p_{0}^R(t),\cdots,p_{N}^R(t))^T$, and $\bm{P^S}(t)$ as the diagonal matrix $diag(1-p_{0}^I(t)-p_{0}^R(t),\cdots,1-p_{N}^I(t)-p_{N}^R(t))$. Then we have
\begin{eqnarray}
% \nonumber to remove numbering (before each equation)
  \bm{p^I}(t+1)&=&\beta_S\bm{P^S}(t)\bm{A}\bm{p^I}(t), \\
  \bm{p^R}(t+1)&=&\bm{p^R}(t)+\bm{p^I}(t).
\end{eqnarray}
Therefore, we obtain the evolution functions for $\bm{p^I}(t)$ and $\bm{p^R}(t)$. Since we select the spreading source randomly, the initial condition is $\bm{p^I}(0)=(1/N,\cdots,1/N)$ and $\bm{p^R}(0)=((N-1)/N,\cdots,(N-1)/N)$. The susceptibility vector $\bm{S}=(S_0,\cdots,S_N)^T=\bm{p^R}(\infty)$. Here $\infty$ indicates the spreading dynamic evolves for a long enough period of time so that no infected individuals are left in the system. In case of sparse networks, the iteration converges very quickly. Therefore the susceptibility can be calculated by this method efficiently.

\begin{figure}
\begin{center}
\includegraphics[width=0.8\columnwidth]{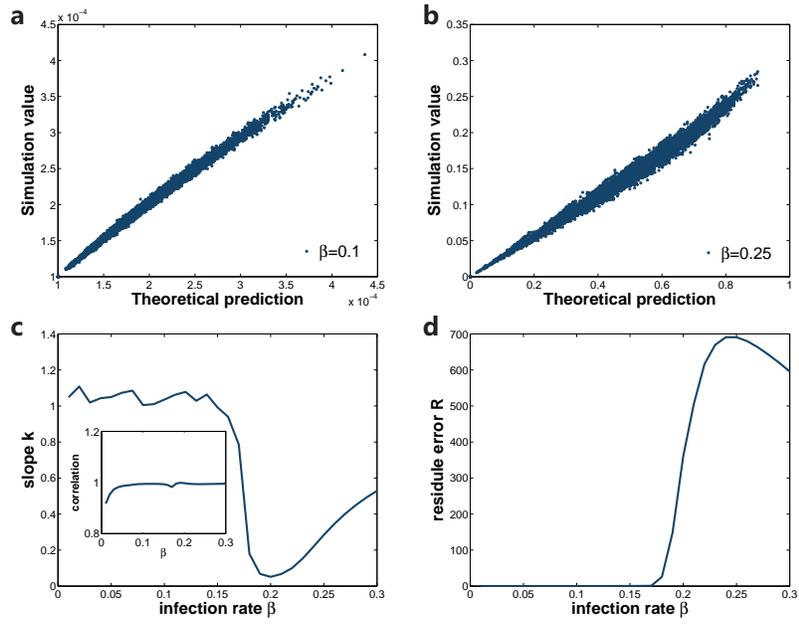} \caption{For ER random network ($N=10^4$, $\langle k\rangle=5$), the simulation values of $S_i$ and theoretical predictions $S^{th}_i$ are displayed in (a) ($\beta_S=0.1$) and (b) ($\beta_S=0.25$). The slope of simulation values versus theoretical predictions for different $\beta_S$ is shown in (c). The inset presents their correlation coefficient. In (d), we show the residual error of theoretical predictions for various $\beta_S$.}
\label{fig3}
\end{center}
\end{figure}

We verify the calculation on Erd\"{o}s-R\'{e}nyi (ER) random networks. We construct the ER random network with size $N=10^4$ and average degree $\langle k\rangle=5$ by randomly assigning $25,000$ links among $10^4$ nodes. Then we perform SIR simulations to obtain simulated susceptibility $S_i$ and calculate the theoretical value $S^{th}_i$ through iterations. For $\beta_S=0.1$, the theoretical values agree well with the simulation results (see Fig.\ref{fig3}(a)). However, for $\beta_S=0.25$, which is above the threshold value $\beta_c=0.2$, theoretical values are considerably larger than simulation values (see Fig.\ref{fig3}(b)).  This discrepancy can be explained by the invalidation of assumptions. For a relatively large infection rate $\beta_S$, the epidemic would occupy a macroscopic fraction of population. Under this condition, the assumption that a node's neighbors are infected independently becomes invalid. Therefore, the exact probability of a node $i$ to be infected by its neighbors at time $t$ should be smaller than $\beta_S p_{i}^S(t)\sum_{j=1}^N a_{ij}p_{j}^S(t)$, which implies our method overestimates nodes' susceptibility. Even though the calculation has the problem of overestimation, the theoretical predictions and simulation values follow a straight line. We fit this line to $S_i=kS^{th}_i$ for different $\beta_S\in[0,0.3]$ and examine the correlation coefficient between $S_i$ and $S^{th}_i$ in the inset of Fig.\ref{fig3}(c). It can be seen that theoretical predictions are highly correlated with simulated susceptibility. However, the slope $k$ varies as $\beta_S$ increases. For $\beta_S<0.15$, the slope $k$ remains around $1$, which means theoretical values are quite accurate. As $\beta_S$ approaching to $\beta_c=0.2$, the slope $k$ decreases dramatically. This implies $S^{th}_i$ becomes larger than $S_i$, even though they still have a linear relationship. This phenomenon is also reflected in Fig.\ref{fig3}(d), in which we display the residual error $R=\sum_i|S_i-S^{th}_i|$ for different $\beta_S$. With $\beta_S$ increasing to $\beta_c$, the residual error $R$ rises abruptly. Therefore, our method predicts susceptibility accurately for infection rate $\beta_S$ smaller than threshold value $\beta_c$, and can be used to rank susceptibility for $\beta_S$ near or larger than $\beta_c$.

\section{Topological indicators for highly susceptible individuals}

Although we have calculated the susceptibility with the standard set of equations of spreading dynamics, it is still convenient to discern highly susceptible individuals with topological indicators in practice. Here we explore some of the most important topological measures in network theory. The most straightforward and simple one is degree $k$, i.e. the number of connections a node maintains. Hubs (high degree nodes) are believed to play significant roles in many dynamical processes. More connections may lead to a higher chance to be infected in an epidemic spreading. However, degree only reflects nodes' number of neighbors. In order to check people's global location in networks, we also examine the $k$-shell index $k_S$ by $k$-shell decomposition analysis \cite{Maksim2010,Pei2013,Seidman1983,Carmi2007,Pei2014,Castellano2014}. The $k$-shell decomposition can be described as follows: firstly we start by recursively removing nodes with degree $k = 1$ until there are only nodes with degree $k > 1$ left in the network. All the nodes removed in this procedure are assigned with $k_S = 1$. In a similar way, we iteratively find out higher $k$-shells until all nodes are pruned. In this procedure, each node will be assigned a unique $k$-shell index $k_S$. Nodes belonging to high (low) $k$-shells are located in the core (periphery) area of the network. Another topological measure, betweenness centrality, quantifies the number of shortest paths passing through each node. In social science, the betweenness centrality of node $i$, denoted by $BC(i)$ is defined as
\begin{equation}
BC(i)=\sum_{s\neq i\neq t}\frac{\sigma_{st}(i)}{\sigma_{st}},
\label{BC}
\end{equation}
where $\sigma_{st}$ is the number of shortest paths between nodes $s$ and $t$, and $\sigma_{st}(i)$ is the number of shortest paths between $s$ and $t$ which pass through node $i$. Consequently, nodes with large betweenness centrality usually hold the vital positions in the shortest pathways between large numbers of nodes pairs. In addition, considering the importance of nodes' neighbors, PageRank, which is originally introduced to rank web pages, is proposed \cite{Brin1998}. The PageRank of a node $i$ in a network can be calculated from
\begin{equation}
P_t(i)=\frac{d}{N}+(1-d)\sum_j\frac{a_{ij}P_{t-1}(j)}{k(j)},
\label{PageRank}
\end{equation}
where $a_{ij}$ is the element of adjacency matrix $\bm{A}$ and $d$ is the jumping probability. In our calculation, as in many previous studies, we set $d=0.15$ conventionally. To sum up, we have chosen four typical topological measures proposed from different perspectives: degree $k$, $k$-shell $k_S$, betweenness centrality $BC$ and PageRank.

\begin{figure}
\begin{center}
\includegraphics[width=0.8\columnwidth]{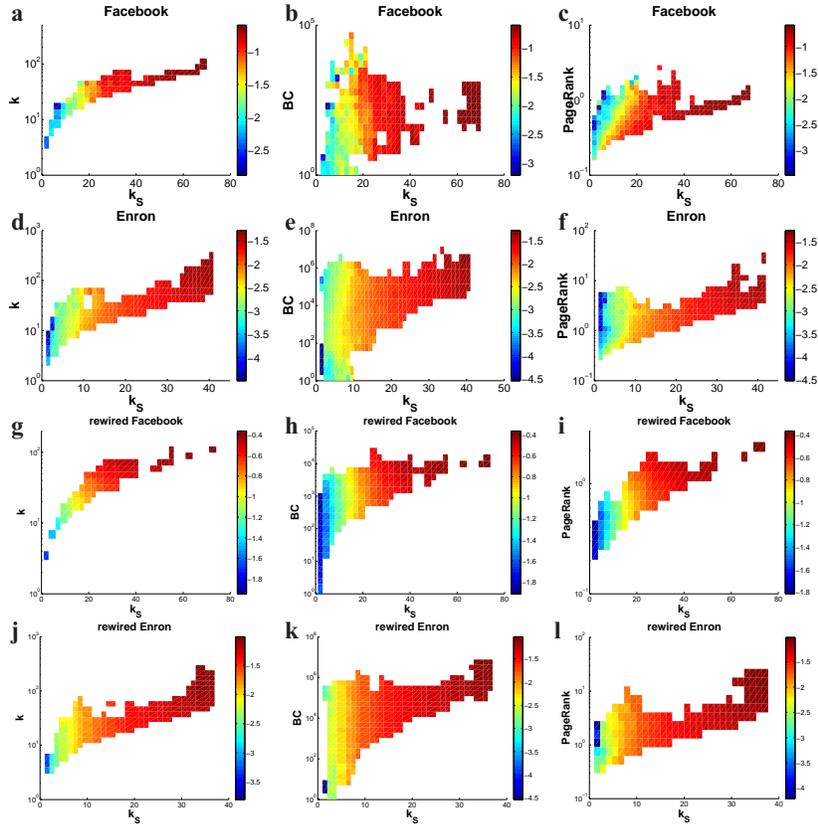} \caption{Comparison of different topological indicators for SIR model. For Facebook (a, b, c), Enron (d, e, f), rewired Facebook (g, h, i) and rewired Enron (j, k, l) networks, we plot the average values of logarithmic susceptibility (base 10) for nodes with combined k-shell $k_S$ and degree $k$ (a, d, g), betweenness centrality $BC$ (b, e, h) and PageRank (c, f, i) respectively. The infection rate $\beta_S$ is $0.03$ and $0.02$ for Facebook and Enron networks. Values are indicated by colors, which are shown in color bars.}
\label{fig4}
\end{center}
\end{figure}

First, we compare the performance of $k$-shell and degree for SIR model. For individuals with a given combination of $k$-shell and degree $(k_S,k)$, we check the average susceptibility $S(k_S,k)$:
\begin{equation}
S(k_S,k)=\sum_{i\in\Upsilon(k_s,k)}\frac{S_i}{N(k_S,k)}.
\label{S(ks k)}
\end{equation}
Here $\Upsilon(k_s,k)$ is the collection of all nodes with $k$-shell and degree $(k_S,k)$, and $N(k_S,k)$ is the number of nodes in the set $\Upsilon(k_s,k)$. The susceptibility $S_i$ of node $i$ is calculated by averaging over $10^4$ SIR realizations in each social network. We display the logarithmic value (base 10) of $S(k_S,k)$ in Fig.\ref{fig4}(a), (d), (g) and (j) for Facebook, Enron, rewired Facebook and rewired Enron social networks respectively. It is observed that most highly susceptible individuals occupy high $k$-shells. For nodes with a fixed degree $k$, the susceptibility can be either large or small, depending on their $k$-shell values. Meanwhile, the individuals located in the same $k_S$ shell have similar susceptibility. This implies, $k$-shell index could better reflect nodes' susceptibility than degree $k$. We also perform same analyses for betweenness centrality $BC$ (see Fig.\ref{fig4}(b),(e),(h),(k)) and PageRank (see Fig.\ref{fig4}(c),(f),(i),(l)). Consistently, $k$-shell index $k_S$ outperforms betweenness centrality and PageRank in predicting the average susceptibility of individuals.

We should also note that our results are not affected by the community structure. We compare our results in the original and rewired networks in Fig.\ref{fig4}. The conclusion is not sensitive to the existence of community structure. No matter where the node locates in the network (between communities or within communities), it is susceptible to SIR spreading processes as long as it has a high $k$-shell index.

\begin{figure}
\begin{center}
\includegraphics[width=0.8\columnwidth]{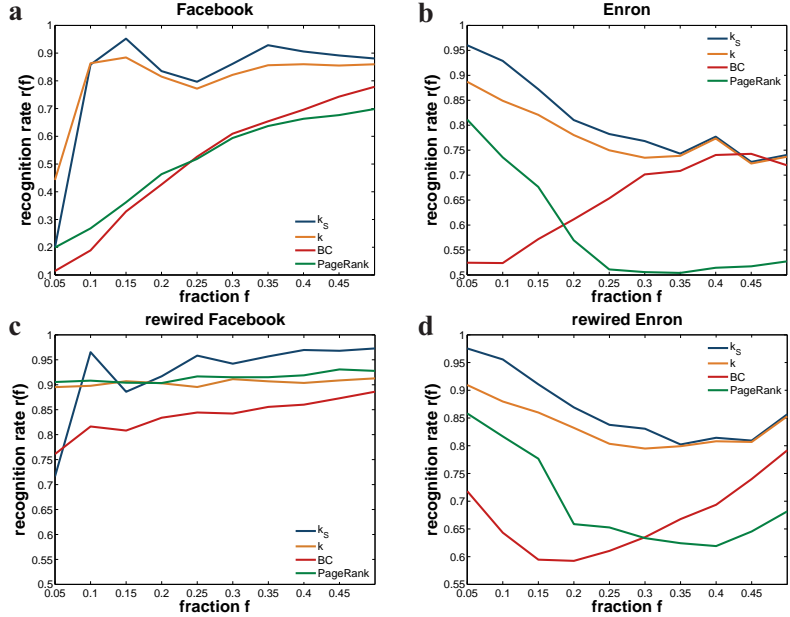} \caption{The recognition rate $r(f)$ for different fraction $f$ in SIR model. For Facebook (a), Enron (b), rewired Facebook (c) and rewired Enron (d) social networks, we display the recognition rate curves $r(f)$ for k-shell $k_S$, degree $k$, betweenness centrality $BC$ and PageRank. The fraction $f$ ranges from 0.05 to 0.5. The infection rate $\beta_S$ in SIR simulations is $0.03$ and $0.02$ for Facebook and Enron networks. }
\label{fig5}
\end{center}
\end{figure}

Although $k$-shell can better predict the average susceptibility of individuals, since the susceptibility for nodes with same topological measures has fluctuations, it is still desirable to explore their accuracy in locating highly susceptible persons. To this end, we use the recognition rate $r(f)$ to quantify the efficiency in identifying individuals of high susceptibility \cite{Pei2014}. The recognition rate $r(f)$ is defined as
\begin{equation}
r(f)=\frac{|S_f\bigcap M_f|}{|S_f|},
\label{r}
\end{equation}
where $S_f$ and $M_f$ are the sets of nodes ranking in top $f$ percentage by susceptibility and topological measures, and $|\cdot|$ is the number of elements in the set. In fact, recognition rate $r(f)$ indicates the proportion of highly susceptible individuals which can be predicted by each measure. The higher $r(f)$ is, the more accurate the predictor is. In Fig.\ref{fig5}, we display the recognition rate $r(f)$ ($f\in[0.05,0.5]$) for Facebook, Enron, rewired Facebook and rewired Enron networks. In most cases, $k$-shell index $k_S$ and degree $k$ outperform betweenness centrality $BC$ and PageRank. And $k_S$ performs better than degree $k$ in general, except when the fraction $f$ is relatively small for Facebook and rewired Facebook networks. This can be explained by the fact that in these networks, nodes with extremely large degree tend to locate in the core region. Therefore, if we pick high-degree nodes, they will also have high $k_S$. However, the number of such hubs is not too large. As we increase the fraction $f$, $k_S$ could still identify more highly susceptible individuals. This phenomenon depends on the exact topological structure of networks. For instance, for tree-like networks, $k_S$ only contains a few values. Under this condition, degree $k$ can clearly better distinguish highly susceptible individuals. Whereas, such simple topological structure usually does not appear in large scale real-world social networks. On the contrary, realistic social networks posses rather complex structure, in which hubs are not necessarily located in the dense core. So in real scenario, $k_S$ is still an effective predictor for highly susceptible individuals.

\begin{figure}
\begin{center}
\includegraphics[width=0.8\columnwidth]{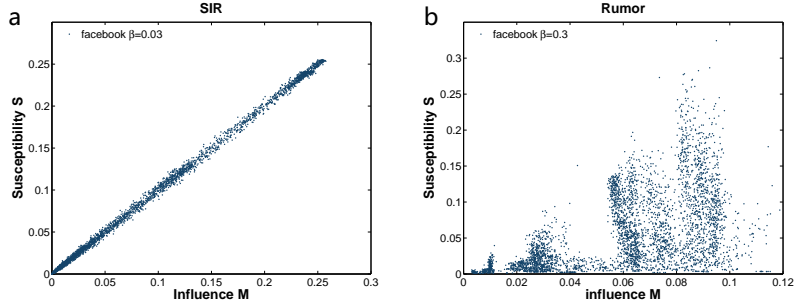} \caption{The relationship between susceptibility $S$ and influence $M$ of SIR model is presented in (a). The infection rate $\beta_S$ is set to be $0.03$. We also display susceptibility $S$ versus influence $M$ of rumor spreading model in (b). $\beta_R$ is set to be $0.3$.}
\label{fig6}
\end{center}
\end{figure}

As pointed out in the supporting information of Ref.\cite{Maksim2010}, for SIR model, nodes' influence and infected probability are highly correlated. We present the relationship between susceptibility $S$ and influence $M$ of SIR model in Fig.\ref{fig6}(a). Influence $M_i$ is defined as the average size of the population infected in a spreading process originating at node $i$. Since $k$-shell outperforms other topological measures in identifying influential spreaders for SIR model \cite{Maksim2010,Pei2014}, it should be also effective in locating highly susceptible individuals. This explains the efficacy of $k$-shell for SIR model. In fact, for any independent-interaction spreading models in which susceptibility and influence are closely correlated, such as SIR and SIS model, $k$-shell should be a good topological indicator for highly susceptible people.

\begin{figure}
\begin{center}
\includegraphics[width=0.8\columnwidth]{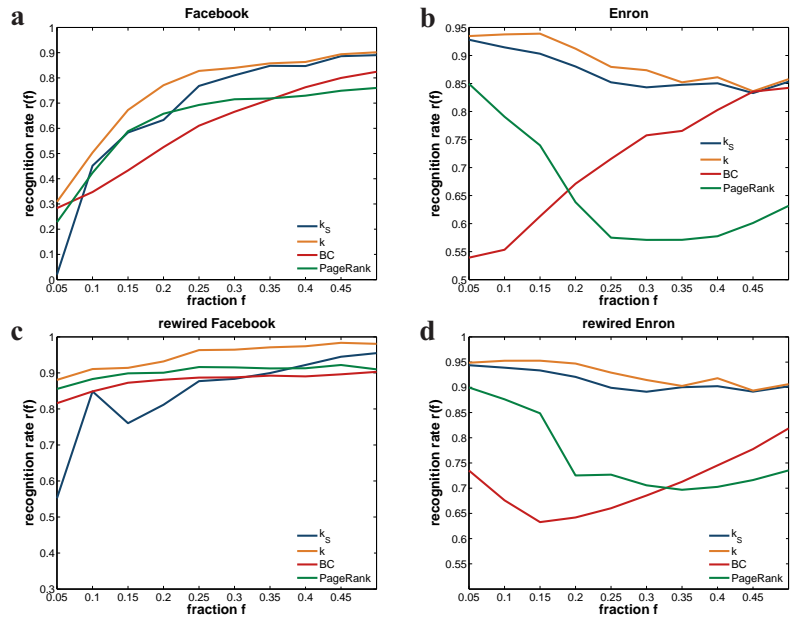} \caption{The recognition rate $r(f)$ for different fraction $f$ in rumor spreading model. For Facebook (a), Enron (b), rewired Facebook (c) and rewired Enron (d) social networks, we display the recognition rate curves $r(f)$ for k-shell $k_S$, degree $k$, betweenness centrality $BC$ and PageRank. The fraction $f$ ranges from 0.05 to 0.5. The infection rate $\beta_R$ in rumor spreading simulations is $0.3$ for all networks. }
\label{fig7}
\end{center}
\end{figure}

However, for rumor spreading model, we find the susceptibility and influence have no clear correlation, as shown in Fig.\ref{fig6}(b). In rumor spreading model, influential spreaders are not necessarily highly susceptible to spreading processes. It has been found that $k$-shell cannot represent nodes' spreading capability in rumor spreading model \cite{Borge2012}. Then it is desirable to find the topological indicator of susceptibility in rumor spreading model. Since the figure like Fig.\ref{fig4} cannot distinguish the performance of some measures clearly, we directly display the recognition rate for each topological indicator in Fig.\ref{fig7}. For all considered networks, degree is capable of recognizing more high-risk people, which differs from the better performance of $k$-shell for SIR model. Also, the community structure will not affect the prominent performance of degree. Whereas, after rewiring links, the recognition rates for all indicators are enhanced. This implies the existence of community structure undermines the effectiveness of topological indicators.

In fact, the different results obtained in SIR model and rumor spreading model can be explained by the spreading mechanisms. In SIR model, the superior performance of $k$-shell relies on the cascading behavior of spreading processes. Starting from the source, the epidemic will diffuse from person to person contagiously and may eventually spread through the majority of population in a ``viral'' way. Usually, a spreading process will last for many generations before it dies out. For nodes located in the core region, not only their neighbors are prone to be infected, but also their neighbors of neighbors have large chances to be infected, and so forth. Consequently, $k$-shell can better describe nodes' susceptibility in SIR model. On the contrary, in rumor spreading model, once a spreader encounters a spreader or stifler neighbor, it will stop spreading the information. This dramatically restrains the cascading behavior. It has been found that nodes with high $k$-shell values serve as ``firewalls'' of rumor spreading \cite{Borge2012}, i.e., once these nodes are infected, they will soon turn into stiflers. Under this situation, high $k$-shell nodes will have lower chances to be infected due to their ``firewall'' neighbors. Since the cascading behavior is restricted, it is sufficient to consider only one-step neighbors. Therefore, degree outperforms $k$-shell in rumor spreading model.

\section{Conclusion and Discussion}

In this paper, we investigate individual's susceptibility in spreading processes. We first study the impact of community structure on people's susceptibility. We find that, although the community structure can reduce the infected population and slow down the spreading processes, it will not affect nodes' susceptibility significantly. For SIR model, we find the topological indicator $k$-shell can better identify highly susceptible individuals, outperforming degree, betweenness centrality and PageRank. This can be explained by the high correlation between nodes' susceptibility and influence. On the contrary, for rumor spreading model, nodes' susceptibility and influence have no clear correlation. Degree performs the best among considered topological measures. We give an explanation of the different results of SIR and rumor spreading models based on spreading mechanisms. In general, for spreading processes that can trigger large-scale cascading behaviors, such as SIR model, nodes in the core region should be more susceptible since they tend to have highly susceptible neighbors, neighbors of neighbors, and so forth. In contrast, in spreading models that rely on short-range diffusion, degree should be a good choice of indicator for susceptibility. Our finding indicates that, when identifying highly susceptible individuals, it is necessary to take both topological features and spreading mechanisms into account.

\section*{Acknowledgments}

%This work is supported by National Natural Science Foundation of China (11290141, 11290143, 11201018), International Cooperation Project No.2010DFR00700, Fundamental Research of Civil Aircraft No. MJ-F-2012-04, Beihang University Innovation and Practice Fund for Graduate and Innovation Foundation of BUAA for PhD Graduates.

%% The Appendices part is started with the command \appendix;
%% appendix sections are then done as normal sections
%% \appendix

%% \section{}
%% \label{}

%% References
%%
%% Following citation commands can be used in the body text:
%% Usage of \cite is as follows:
%%   \cite{key}          ==>>  [#]
%%   \cite[chap. 2]{key} ==>>  [#, chap. 2]
%%   \citet{key}         ==>>  Author [#]

%% References with bibTeX database:

\bibliographystyle{model1a-num-names}
%\bibliography{<your-bib-database>}

\begin{thebibliography}{00}

%% \bibitem must have the following form:
%%   \bibitem{key}...
%%




% \bibitem{}
\bibitem{Keeling2008}
M.J. Keeling, P. Rohani, Modeling Infectious Diseases in Humans and Animals, Princeton University Press, 2008.

\bibitem{Brockmann2013}
D. Brockmann, D. Helbing, Science 342 (2013) 1337.

\bibitem{Maksim2010}
M. Kitsak, L.K. Gallos, S. Havlin et al, Nat. Phys. 6 (2010) 888.

\bibitem{Albert2000}
R. Albert, H. Jeong, A.-L. Barab\'{a}si, Nature 406 (2000) 378.

\bibitem{Reis2014}
S.D.S. Reis, Y. Hu, A. Babino et al, Nat. Phys. 10 (2014) 762.

\bibitem{Pei2012}
S. Pei, S. Tang, S. Yan et al, Phys. Rev. E 86 (2012) 021909.

\bibitem{Boguna2002}
M. Bogun\'{a}, R. Pastor-Satorras, Phys. Rev. E 66 (2002) 047104.

\bibitem{Moreno2004}
Y. Moreno, M. Nekovee, A.F. Pacheco, Phys. Rev. E 69 (2004) 066130.

\bibitem{Borge2012}
J. Borge-Holthoefer, Y. Moreno, Phys. Rev. E 85 (2012) 026116.

\bibitem{Pei2012a}
S. Pei, S. Tang, X. Zhang et al, Physica A 391 (2012) 2023.

\bibitem{Hu2014}
Y. Hu, S. Havlin, H.A. Makse, Phys. Rev. X 4 (2014) 021031.

\bibitem{Teng2014}
X. Teng, S. Yan, S. Tang et al. Physica A 402 (2014) 141.

\bibitem{Pastor2001}
R. Pastor-Satorras, A. Vespignani, Phys. Rev. Lett. 86 (2001) 3200.

\bibitem{Newman2002}
M.E.J. Newman, Phys. Rev. E 66 (2002) 016128.

\bibitem{Yan2013}
S. Yan, S. Tang, S. Pei et al, Physica A 392 (2013) 3846.

\bibitem{Girvan2002}
M. Girvan, M.E.J. Newman, Proc. Natl. Acad. Sci. USA 99 (2002) 7821.

\bibitem{Daniel2011}
D. Grady, C. Thiemann, D. Brockmann, Nat. Comm. 3 (2011) 864.

\bibitem{Onnela2006}
J.-P. Onnela, J. Saram\"aki, J. Hyv\"onen et al, Proc. Natl. Acad. Sci. USA 104 (2006) 7332.

\bibitem{Freeman1979}
L.C. Freeman, Soc. Netw. 1 (1979) 215.

\bibitem{Zeng2013}
A. Zeng, C.J. Zhang, Phys. Lett. A 377 (2013) 1031.

\bibitem{Brin1998}
S. Brin, L. Page, Comput. Networks ISDN 30 (1998) 107.

\bibitem{Lu2011}
L. L\"{u}, Y.C. Zhang, C.H. Yeung et al, PLoS One 6 (2011) e21202.

\bibitem{Leskovec2012}
J. Leskovec, J.J. Mcauley, Advances in neural information processing systems (2012) 539.

\bibitem{Klimt2004}
B. Klimt, Y. Yang, Machine Learning (2004) 217.

\bibitem{Blondel2008}
V.D. Blondel, J.-L. Guillaume, R. Lambiotte et al, J. Stat. Mech 10 (2008) P10008.

\bibitem{Hethcote2000}
H.W. Hethcote,SIAM Rev. 42 (2000) 599.

\bibitem{Yan2014}
S. Yan, S. Tang, S. Pei et al, Phys. Rev. E 90 (2014) 022808.

\bibitem{Centola2010}
D. Centola, Science 329 (2010) 1194.

\bibitem{Centola2011}
D. Centola, Science 334 (2011) 1269.

\bibitem{Karsai2014}
M. Karsai, N. Perra, A. Vespignani, Sci. Rep. 4 (2014) 4001.

\bibitem{Muchnik2013}
L. Muchnik, S. Pei, L.C. Parra et al, Sci. Rep. 3 (2013) 1783.

\bibitem{Li2014}
W. Li, S. Tang, S. Pei et al, Physica A 397 (2014) 121.

\bibitem{Weng2013}
L. Weng, F. Menczer, Y.Y. Ahn, Sci. Rep. 3 (2013) 2522.

\bibitem{Smilkov2014}
D. Smilkov, C.A. Hidalgo, L. Kocarev, Sci. Rep. 4 (2014) 4795.

\bibitem{Pei2013}
S. Pei, H.A. Makse, J. Stat. Mech. 12 (2013) P12002.

\bibitem{Seidman1983}
S.B. Seidman, Soc. Netw. 5 (1983) 269.

\bibitem{Carmi2007}
S. Carmi, S. Havlin, S. Kirkpatrick et al, Proc. Natl. Acad. Sci. USA 104 (2007) 1115.

\bibitem{Pei2014}
S. Pei, L. Muchnik, J.S. Andrade Jr et al, Sci. Rep. 4 (2014) 5547.

\bibitem{Castellano2014}
C. Castellano, R. Pastor-Satorras, Sci. Rep. 2 (2012) 371.

\bibitem{Brandes2001}
U. Brandes, J. Math. Sociol. 25 (2001) 163.




\end{thebibliography}

%% Authors are advised to submit their bibtex database files. They are
%% requested to list a bibtex style file in the manuscript if they do
%% not want to use model1a-num-names.bst.

%% References without bibTeX database:

\end{document}